\documentclass[prl,twocolumn,amsmath,amssymb]{revtex4} \usepackage{graphicx,dcolumn,bm,color} 

\begin{document} 

\title{Hidden Structural Order in Orthorhombic Ta$_2$O$_5$ }
\author{Sung-Hoon~Lee} \email{sung-hoon.lee@samsung.com}
\author{Jongseob~Kim} \email{jongs.kim@samsung.com}
\author{Sae-Jin~Kim}
\author{Sungjin~Kim} 
\author{Gyeong-Su~Park} 
\affiliation{Samsung Advanced Institute of Technology, Yongin 446-712, Korea}

\begin{abstract}
{ 

We investigate using first-principles calculations the atomic structure of the orthorhombic phase of Ta$_2$O$_5$. Although this structure has been studied for decades, the correct structural model is controversial owing to the complication of structural disorder. We identify a new low-energy high-symmetry structural model where all Ta and O atoms have correct formal oxidation states of $+5$ and $-2$, respectively, and the experimentally reported triangular lattice symmetry of the Ta sublattice appears dynamically at finite temperatures. To understand the complex atomic structure of the Ta$_2$O$_3$ plane, a triangular graph-paper representation is devised and used alongside oxidation state analysis to reveal infinite variations of the low-energy structural model. The structural disorder of Ta$_2$O$_5$ observed in experiments is attributed to the intrinsic structural variations, and oxygen vacancies that drive collective relaxation of the O sublattice.

}

\end{abstract}

\maketitle

Tantalum pentoxide (Ta$_2$O$_5$) is one of the most extensively studied transition metal oxides due to its potential for technological applications such as anti-reflection coatings, photocatalysis, and high-k dielectrics for high-density transistors \cite{Kingon00,Takahara01,Chaneliere98}. 
It has recently attracted interest as the material of choice for resistance-change memory and the related memristor \cite{Lee11,Yang10b,Miao11AM}, however, its crystalline structure, which is critical to its unique structural and electronic properties, is still unclear.

Ta$_2$O$_5$ has an orthorhombic phase up to $\sim 1350^\circ$C, above which a transition to a tetragonal phase occurs \cite{Lagergren52,Garg96}. 
Although the low-temperature orthorhombic phase is of technological relevance, it has proven difficult to characterize structurally.
Early X-ray diffraction studies \cite{Calvert62,Lehovec64} found that the orthorhombic Ta$_2$O$_5$ phase consists of two-dimensional (2D) layers of Ta$_2$O$_3$ in the \textit{ab} plane, and two-fold coordinated O atoms that connect the Ta atoms in adjacent layers (forming linear chains of -Ta-O-Ta-O-Ta-) along the \textit{c} direction. 
In the Ta$_2$O$_3$ plane, it was understood that the Ta atoms form a triangular arrangement, but the arrangement of O atoms was unresolved, with many weak superstructure lines \cite{Lehovec64}. 
Later, Roth and coworkers reported that the Ta$_2$O$_3$ plane consists of a disordered array of Ta atoms in octahedral and pentagonal bipyramidal arrangements, with an 11-fold increase of the lattice parameters in the \textit{b} direction \cite{Stephenson71V}. 
The superstructure periodicity was shown to change when aliovalent cations, such as W or Li, were incorporated in small amounts to stabilize the phase \cite{Stephenson71II,Grey05}.
The superstructures consist of a set of small-unit basic structures and invoked the concept of infinitely adaptive structures \cite{Anderson73JCS} for nonstoichiometric Ta$_2$O$_{5-x}$ that---within certain composition limits---every composition orders into a different superstructure.

Extensive theoretical studies have been made with regards to the bulk properties, dielectric response, and oxygen vacancies (and their diffusion) of Ta$_2$O$_5$ \cite{Fukumoto97,Sahu04,Gu09,Wu11,Clima10,Ivanov11,Andreoni10,Sawada99,Ramprasad03JAP}.
In these studies, the low-temperature Ta$_2$O$_5$ phase was typically modeled by high-symmetry structural models, such as the orthorhombic $\beta$ model \cite{Aleshina02}, or the hexagonal $\delta$ model \cite{Fukumoto97} (See Fig.~1). 
These models assume a triangular sublattice of Ta, as shown in early experiments \cite{Calvert62,Lehovec64}, with one or two Ta$_2$O$_5$ units per unit cell.
Recently, however, Andreoni and Pignedoli found that both the $\beta$ and $\delta$ models have instability modes with imaginary vibration frequencies \cite{Andreoni10}. 
This instability highlights the need to clearly identify the atomic structure of the orthorhombic Ta$_2$O$_5$ phase in order to improve fundamental understanding of the material, and also to accelerate technological applications.
An attempt to pinpoint the origin of the structural disorder \cite{Lehovec64,Stephenson71II,Stephenson71V,Grey05,Anderson73JCS} should also be made.

In this Letter, we use first-principles calculations to determine the key structural motif of the orthorhombic phase of Ta$_2$O$_5$ and the origin of structural disorder.  
We propose a new structural model that consists of two Ta$_2$O$_5$ units per unit cell, is stable, and accounts very well for experimentally observed structural and electronic properties. 
Our first-principles molecular dynamics (MD) simulations show that the Ta atoms exhibit a triangular lattice symmetry at finite temperatures that is otherwise hidden in zero-temperature calculations.
This hidden structural order suggests that the complex 2D atomic structure in the Ta$_2$O$_3$ plane can be mapped one-to-one onto triangular graph-paper.
In this graphical representation, the low-energy atomic configurations can be deduced by simply counting the formal oxidation states of Ta and O atoms, and turn out to be not unique but of infinite variations with the same structural motif.
Our findings indicate that the structural disorder, which has hindered experimental determination of the structure, is an intrinsic property of the orthorhombic Ta$_2$O$_5$ phase.
 We also investigate the impact of oxygen vacancies on the structural disorder.

\begin{figure}[t] \includegraphics[width=8.7cm]{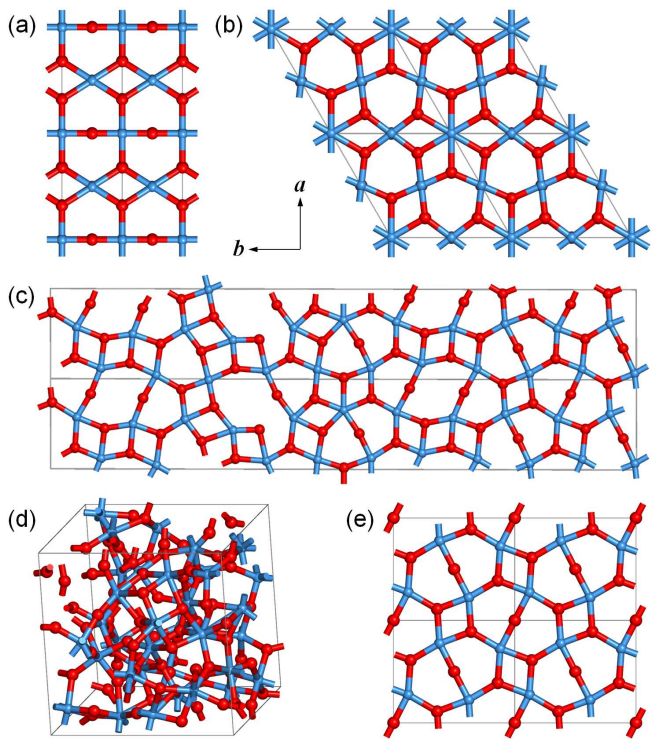} \caption{
(color online).
Structural models for low-temperature Ta$_2$O$_5$. The 2D atomic structures in the Ta$_2$O$_3$ plane are shown, where the blue and red spheres represent Ta and O atoms, respectively. (a) The orthorhombic $\beta$ model (space group \textit{Pmmm}) \cite{Aleshina02}. (b) The hexagonal $\delta$ model (space group \textit{P6/mmm}) \cite{Fukumoto97}. (c) The relaxed atomic structure of the 11 f.u.\ model \cite{Stephenson71V}. (d) An amorphous model. (e) The orthorhombic $\lambda$ model (space group \textit{Pbam}).
} \end{figure}

The first-principles investigation presented here is based on density functional calculations that employ the generalized gradient approximation (GGA) \cite{PBE} and the projector-augmented-wave method, as implemented in VASP \cite{Kresse96,Kresse99}. Valence electron wavefunctions were expanded in a planewave basis set with a cutoff energy of 400 eV. The k-point integration was performed using a uniform k-point mesh with a mesh spacing $< 0.3$ \AA$^{-1}$. The equilibrium lattice parameters are obtained by performing the variable cell optimization that is implemented in VASP. We confirmed that the relative energetic stability between different structural models---with respect to the cutoff energy, k-point sampling, and volume relaxation---converged well below 0.1 eV per Ta$_2$O$_5$ unit.

The crystal structure of the orthorhombic Ta$_2$O$_5$ phase consists of 2D layers of Ta$_2$O$_3$ that are connected to each other by two-fold coordinated O atoms between Ta atoms in adjacent layers \cite{Calvert62,Lehovec64}.
Several structural models have been suggested, where the atomic arrangements in the Ta$_2$O$_3$ plane differ.
Figure 1 shows representative structural models for low-temperature Ta$_2$O$_5$: orthorhombic $\beta$, hexagonal $\delta$, and 11 formular unit (f.u.) models. 
The 11 f.u.\ model [Fig.~1(c)] is structurally relaxed within the present calculations.
For comparison, we generated amorphous structures [Fig.~1(d)] by performing first-principles MD simulations for a cubic cell containing 17 Ta$_2$O$_5$ units and of the volume given by the experimental mass density of 8.2 g/cm$^{-3}$.

\begin{table}[t]
\caption{\label{tab:table1} Calculated lattice parameters and total energies of low-temperature Ta$_2$O$_5$.
The experimental lattice parameters are also listed for comparison.
Total energies ($\Delta E$) are given in eV/f.u.\ with respect to an amorphous structure. }
\begin{ruledtabular}
\begin{tabular}{cccccc}
& \multicolumn{3}{c}{Lattice param. (\AA)} & \\
Model & \textit{a} & \textit{b} & \textit{c} & $\Delta E$ \\
\hline
Expt.~\cite{Calvert62} & 6.18 & 3.66 & 3.88 & \\ 
Expt.~\cite{Lehovec64} & 6.20 & 3.66 & 3.89 & \\ 
Expt.~\cite{Terao67} & 6.18 & 3.66 & 3.88 & \\
Expt.~\cite{Stephenson71V} & 6.20 & $11\times3.66$ & 3.89 & \\
Expt.~\cite{Aleshina02} & 6.22 & 3.68 & 3.90 & \\ 
Expt.~\cite{Grey05}  & 6.19 & $19\times3.66$ & 3.89 & \\ 
Expt.~\cite{Terao67}\footnote{Hexagonal phase.} & \multicolumn{2}{c}{3.62} & 3.88 & \\
$\beta$ model & 6.52 & 3.69 & 3.89 & $+1.27$ \\
$\delta$ model$^a$ & \multicolumn{2}{c}{$2\times3.67$} & 3.89 & $+0.91$ \\
11 f.u. & 6.34 & $11\times3.73$ & 3.85 & $-0.75$ \\
$\lambda$ model & 6.25 & $2\times3.70$ & 3.83 & $-0.81$ \\
\end{tabular}
\end{ruledtabular}
\end{table}

Table~I shows the calculated lattice parameters and relative total energies of the structural models given in Fig.~1. 
The experimental lattice parameters are all consistent with each other regardless of their disparate lattice parameters in the \textit{b} direction; $a = 6.20$ \AA, $b=3.66$ \AA, and $c=3.88$ \AA.
The calculated lattice parameters for the $\beta$ and $\delta$ models are consistent with the experimentally determined values of the orthorhombic and hexagonal phases, respectively, as was found in previous theoretical works \cite{Fukumoto97,Sahu04,Gu09,Clima10,Wu11,Ivanov11}, and regarded as evidence that these models are correct.
Our calculated total energies, however, show that both models are energetically unstable, with total energies of $+1.27$ and $+0.91$ eV/f.u., respectively, with respect to the amorphous structure.
An energy difference of $+1$ eV/f.u.\ is large enough to rule out these models.
The recent phonon mode analysis of the $\beta$ and $\delta$ models that found vibration modes with imaginary frequencies \cite{Andreoni10} corroborates the present result.

On the other hand, we found that the disordered 11 f.u.\ model \cite{Stephenson71V} is energetically stable after structural relaxation; its total energy is $-0.75$ eV/f.u.\ with respect to the amorphous structure \cite{MoreOn11fuModel}. 
It is likely that a minimized structural model of the orthorhombic Ta$_2$O$_5$ phase exists that is stable and free of disorder and thus would be advantageous to understand key structural and energetic features of the system.

We propose a new low-energy high-symmetry structural model, $\lambda$, with two formula units per cell [Fig.~1(e)]. 
All the Ta atoms in this model have the same four-fold in-plane coordination with three three-fold coordinated O atoms and one two-fold coordinated O atom, satisfying space group symmetry of \textit{Pbam}. 
The calculated atomic positions are given in Supplemental Material \cite{SuppMater}.
The calculated lattice parameters are in good agreement with the experimental values of orthorhombic Ta$_2$O$_5$, as shown in Table~I, and the model is energetically stable; its total energy is $-0.06$ eV/f.u.\ with respect to the 11 f.u.\ model.

The $\lambda$ model also accounts very well for the electronic properties of Ta$_2$O$_5$. 
We calculated the density of states (DOS) of the four structural models discussed so far ($\beta$, $\delta$, amorphous, and $\lambda$), by performing hybrid functional calculations \cite{Heyd03} for the GGA-optimized structures to correct for the band gap underestimation of the GGA calculations.
Whereas the $\beta$ and $\delta$ models have band gaps of $\sim 2$ eV, the amorphous structure and the $\lambda$ model have band gaps of $\sim 4$ eV (see Supplemental Material \cite{SuppMater}), which is in agreement with the experimental band gap of 4.0 eV \cite{Fleming00}.

Interestingly, the $\lambda$ model does not exhibit clearly the triangular lattice symmetry of the Ta sublattice, which has been shown in experiments \cite{Lehovec64}.
We thus examined the dynamic effects of the $\lambda$ model by performing first-principles constant-volume MD simulations for a $3\times3\times1$ supercell at 1000 K for 5 ps, based on the calculated equilibrium lattice constants.
The atomic trajectories in the Ta$_2$O$_3$ plane [Fig.~2(a)] shows that the Ta atoms form a well-ordered triangular lattice,
indicating that the triangular lattice symmetry of the Ta sublattice appears in the time-averaged sense. 
In the case of the O atoms, the three-fold coordinated O atoms are relatively well localized in their equilibrium positions. On the other hand, the two-fold coordinated O atoms move widely in a lateral direction and temporarily form three-fold coordination, as seen in the snapshot [Fig.~2(b)].

The time-averaged atomic configurations suggest a new viewpoint from which to understand the atomic structure of the low-temperature Ta$_2$O$_5$.
In Fig.~2(c) to (e), we have redrawn the $\lambda$, $\beta$, and $\delta$ models onto triangular graph-paper.
The Ta atoms, marked as blue points, are regularly placed in a triangular lattice with a periodicity of $(\sqrt3\times\sqrt3)R30^\circ$.
The three-fold coordinated O atoms are represented as red points at vertices, and the two-fold coordinated O atoms moving between two three-fold coordination sites are represented as solid red lines.
As shown in Fig.~2, the 2D atomic structures of all of the $\lambda$, $\beta$, and $\delta$ models can be perfectly described as points and lines on triangular graph-paper. 

\begin{figure}[t] \includegraphics[width=8.7cm]{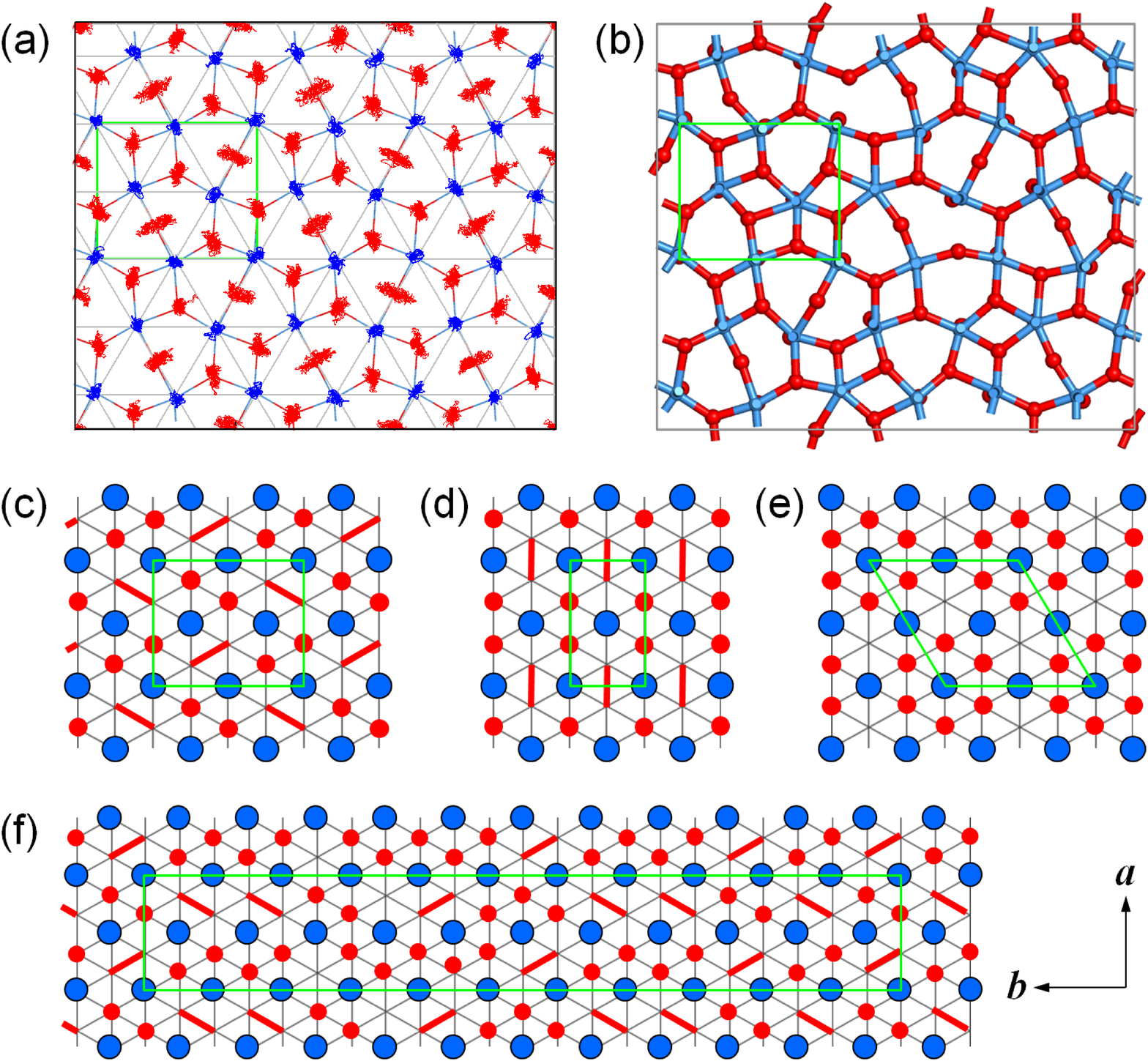} \caption{
(color online).
(a) Atomic trajectories in the Ta$_2$O$_3$ plane obtained from MD simulations where Ta and O atoms are represented by blue and red, respectively.
(b) A snapshot of the MD dynamics.
(c--f) Schematic representations of the 2D atomic structures in the Ta$_2$O$_3$ plane of the $\lambda$, $\beta$, $\delta$, and 11 f.u.\ models, respectively, on triangular graph-paper. Unit cells are marked with green lines.
} \end{figure}

From this graph-paper representation, the low-energy atomic configurations can be easily deduced by counting the oxidation states of individual Ta and O atoms. Because Ta$_2$O$_5$ forms ionic bonds, the formal oxidation states of Ta and O in the ground-state configuration are $+5$ and $-2$, respectively.
By assigning the oxidation state of $-2$ to all O atoms, the oxidation states of Ta atoms can be counted by simply assuming that two-fold coordinated O atoms take a single electron, and three-fold coordinated O atoms take $2/3$ of an electron from each bonded Ta atom.

In the $\lambda$ model, all of the Ta atoms make bonds with three three-fold coordinated O atoms and one two-fold coordinated O atom in the Ta$_2$O$_3$ plane, and two two-fold coordinated O atoms out of the plane.
Thus, they have the correct oxidation states of $+5$. 
On the other hand, in the $\beta$ model, one half of the Ta atoms make bonds with four three-fold coordinated O atoms, and the other half make bonds with two three-fold coordinated and two two-fold coordinated O atoms in the plane.
This results in oxidation states of $+4\frac{2}{3}$ and $+5\frac{1}{3}$, respectively.
Similarly, the Ta atoms in the $\delta$ model have oxidation states of $+4\frac{2}{3}$ and $+6$. 
The present analysis of oxidation states accounts very well for the energetic stability of different models, since a deviation from a value of $+5$ for Ta atoms indicates partial loss of the electrostatic energy gain. 
Thus, the fact that only the $\lambda$ model has the correct formal oxidation states is a clue to its stable state.
The improper oxidation states of the Ta atoms in the $\beta$ and $\delta$ models can be held responsible for their small band gaps; the empty subgap states come from less oxidized Ta atoms due to reduced Coulomb repulsion from neighboring O ions.

This simple analysis can also be used to explain the high stability of the relaxed 11 f.u.\ model. 
As shown in Fig.~2(e), 16 of the 22 Ta atoms per unit cell have oxidation states of $+5$, whereas the other 6 atoms have oxidation states of $+4\frac{2}{3}$ or $+5\frac{2}{3}$.  Thus, the 11 f.u.\ model can be understood as a disordered variation of the $\lambda$ model with more than 70\% of Ta atoms in the same structural motif as the $\lambda$ model. In fact, in the original 11 f.u.\ model \cite{Stephenson71V}, only 55\% of Ta atoms had the correct oxidation state; by imposing structural relaxation in our calculations the model was forced to exhibit more stable motifs.

Using the graph-paper representation, we investigated the possibility of additional low-energy atomic configurations.
For all Ta atoms to have the correct oxidation state of $+5$, we searched for atomic configurations having the structural motif of the $\lambda$ model, and found that there can be infinitely many variations of the $\lambda$ model.
Configurations with 1--4 f.u.\ per unit cell are shown in Fig.~3(a). 
Our first-principles calculations showed that, except for the configuration with 1 f.u.\ per unit cell and the configurations elongated in the \textit{a} direction, which experience high strain, the newly found configurations have similar energies to the $\lambda$ model (within $\pm 0.1$ eV/f.u.).
This result suggests that in the thermodynamic ground state the O sublattice would arrange itself locally in a similar way to the $\lambda$ model, but would be globally disordered, indicating that structural disorder is an intrinsic property of the orthorhombic Ta$_2$O$_5$ phase.

Structural disorder can also be induced by extrinsic factors---such as oxygen vacancy defects---in this system. 
In the $\lambda$ model, there are three distinct O sites: the two- and three-fold coordination sites in the Ta$_2$O$_3$ plane; and the two-fold coordination site of the -Ta-O-Ta- interlayer chain. In O-rich conditions, the calculated formation energies of neutral oxygen vacancy defects are 5.9, 4.4, and 5.6 eV, respectively.
The vacancy formation at the three-fold coordination site is the most favored because a collective structural relaxation in the Ta$_2$O$_3$ plane occurs, as shown in Fig.~3(b). 
When a three-fold coordinated O atom is removed, two neighboring two-fold coordinated O atoms move to form a three-fold coordination, introducing charge imbalance in neighboring Ta atoms and in turn inducing the movement of more O atoms.
Thus, a single oxygen vacancy can distort the O sublattice significantly.

\begin{figure}[t] \includegraphics[width=8.7cm]{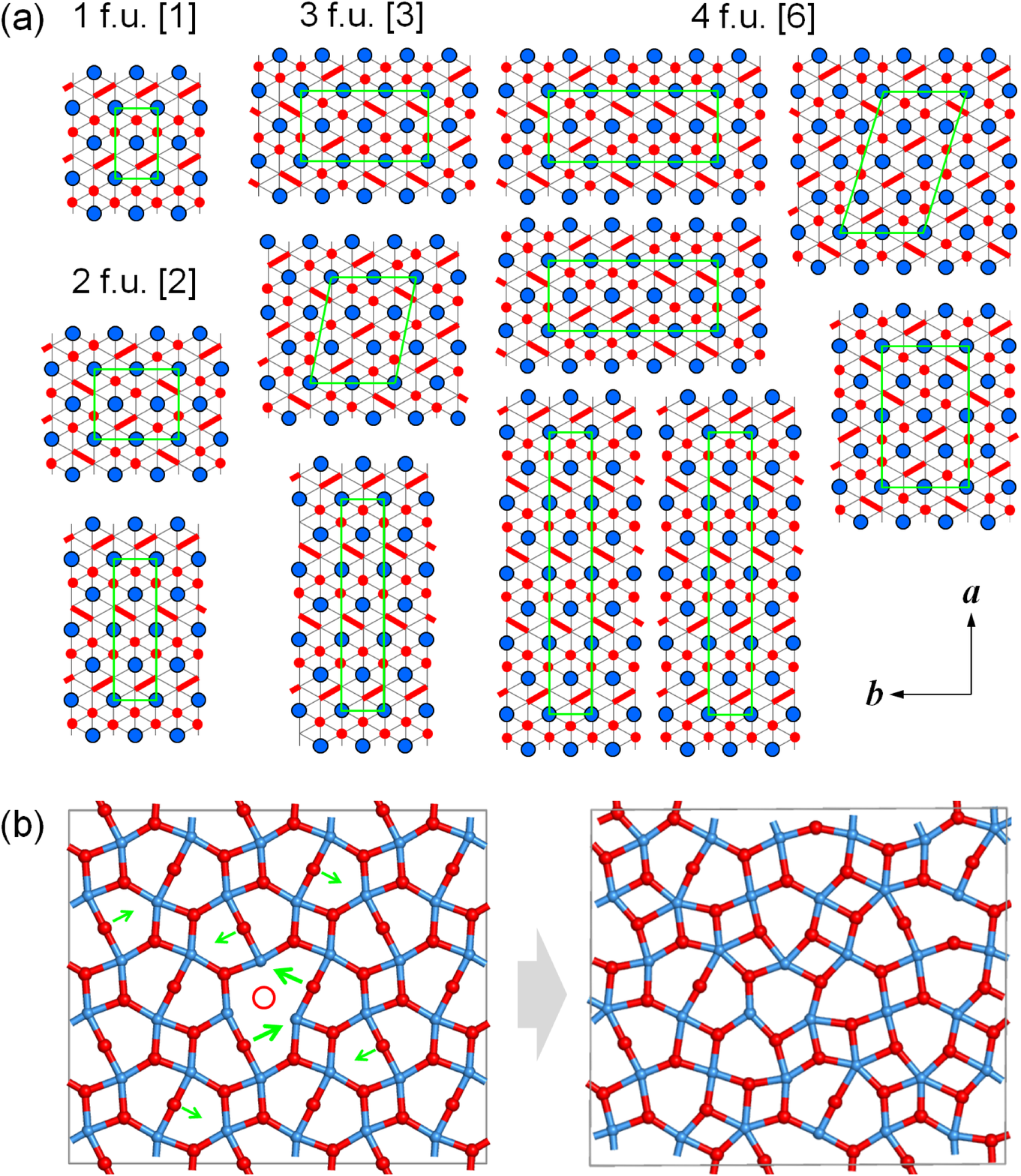} \caption{
(color online).
(a) Atomic configurations with the correct oxidation states for all atoms (i.e., $+5$ for Ta and $-2$ for O). 
Configurations that have 1 to 4 Ta$_2$O$_3$ units per unit cell are shown.
Unit cells are marked with green lines.
(b) Collective relaxation of the O sublattice for the formation of a single oxygen vacancy, before (left) and after (right) the relaxation.
} \end{figure}

There are two points to note here. First, the collective relaxation involves only a small displacement of the O atoms and leaves the arrangement of Ta atoms almost intact. 
As the density of O vacancies increases, the O sublattice will become readily disordered, whereas the triangular arrangement of the Ta sublattice will be preserved to a certain density of O vacancies.
Secondly, although the crystal structure may appear disordered, the local arrangement of O atoms guarantees as much as possible an oxidation state of $+5$ for nearby Ta atoms.
This rearrangement of the O sublattice in the Ta$_2$O$_{5-x}$ phases sheds new light on the old concept of infinitely adaptive structures \cite{Anderson73JCS}.

To conclude, we have presented a new low-energy structural model which incorporates the key structural motif of the orthorhombic phase of Ta$_2$O$_5$.
In the $\lambda$ model, each Ta atom has four in-plane bonds (three three-fold coordinated O atoms and one two-fold coordinated O atoms), with formal oxidation states of the Ta and O atoms of $+5$ and $-2$, respectively.
Based on time-averaged atomic configurations, we revealed that the Ta sublattice forms a triangular arrangement and each two-fold coordinated O atom occupies two three-fold coordination sites. 
We have also clarified the origin of structural disorder in this system using the $\lambda$ model.
The structural disorder of the Ta$_2$O$_{5-x}$ phases was attributed to two factors: intrinsic infinite variations of the $\lambda$ model; and extrinsic collective relaxation of the O sublattice during oxygen vacancy formation. We anticipate that the $\lambda$ model and oxygen vacancy structure presented here will contribute significantly to deeper understanding of the orthorhombic Ta$_2$O$_5$ phase and its utility for technological applications.
Also, we argue that the graph-paper representation with oxidation state analysis could be a useful tool in investigating other layered oxide materials such as uranium oxides. 

We thank Tae-Ho Lee, Kihong Kim, Eunae Cho, Wonjoon Son, and Soogine Chong for helpful discussions.


\begin{thebibliography}{32}
\bibitem{MoreOn11fuModel} {The disordered 11 f.u.\ model has fractional occupancies for four O atoms. We examined a couple of distinct configurations and found that their energetic stability is almost the same, with their energy difference being $0.01$ eV/f.u when relaxed.}
\end{thebibliography}

\end{document}